# Numerical Simulations of Particulate Suspensions via a Discretized Boltzmann Equation Part II. Numerical Results

Anthony J. C. Ladd

*Lawrence Livermore National Laboratory,*

*Livermore, California 94550.*

(June 30, 1993)

## Abstract

A new and very general technique for simulating solid-fluid suspensions has been described in a previous paper (Part I); the most important feature of the new method is that the computational cost scales *linearly* with the number of particles. In this paper (Part II), extensive numerical tests of the method are described; for creeping flows, both with and without Brownian motion, and at finite Reynolds numbers. Hydrodynamic interactions, transport coefficients, and the short-time dynamics of random dispersions of up to 1024 colloidal particles have been simulated.

# I. INTRODUCTION

In a previous paper (Ladd, 1993b), Part I of this series, the theoretical foundations of a numerical method for simulating solid-fluid suspensions was described. By combining Newtonian dynamics of the solid particles with a lattice-Boltzmann model of the fluid, a flexible and efficient algorithm was constructed, whose computational cost scales linearly with the number of particles. In this paper, the focus will be on the implementation of the algorithm for a variety of flow problems, and the numerical results.

In section II, the numerical method described in Part I will be briefly recapped. Simulations of creeping-flow hydrodynamic interactions are described in section III; results are presented for periodic arrays, containing one or two spheres per unit cell, and for random dispersions of spheres. Section IV describes simulations at finite Reynolds number; flows past cylinders and spheres are compared with finite-difference and finite-element computations at Reynolds numbers up to 200. Section V describes time-dependent flows, before the hydrodynamic interactions are fully developed, and section VI deals with fluctuations and Brownian motion. The work is summarized in section VII.

# II. NUMERICAL METHOD

In the lattice-Boltzmann approximation, the state of the fluid system is characterized by the discretized one-particle velocity distribution function $n_i(\mathbf{r}, t)$, which describes the number of particles at a particular node of the lattice $\mathbf{r}$, at a time $t$, with the discrete velocity $\mathbf{c}_i$. The hydrodynamic fields, mass density $\rho$, momentum density $\mathbf{j} = \rho \mathbf{u}$, and momentum flux $\mathbf{\Pi}$, are moments of this velocity distribution:

$$\rho = \sum_i n_i, \quad \mathbf{j} = \sum_i n_i \mathbf{c}_i, \quad \mathbf{\Pi} = \sum_i n_i \mathbf{c}_i \mathbf{c}_i. \tag{2.1}$$

The specific model used in this work has 18 different velocities corresponding to the near-neighbor and second-neighbor directions of a simple cubic lattice; a complete list of the velocities is given in Table I. All quantities in this paper are expressed in "lattice units", for which the distance between nearest-neighbor nodes and the time for the particles to travel from node to node are both unity.

The time evolution of the distribution functions $n_i$ is described by a discrete analogue of the Boltzmann equation (Frisch et al., 1987),

$$n_i(\mathbf{r} + \mathbf{c}_i, t + 1) = n_i(\mathbf{r}, t) + \Delta_i(\mathbf{r}, t), \tag{2.2}$$

where $\Delta_i$ is the change in $n_i$ due to instantaneous molecular collisions at the lattice nodes. The post-collision distribution $n_i + \Delta_i$ is propagated to the neighboring node lying in the direction of the velocity vector $\mathbf{c}_i$. A detailed description of the collision process is given in Part I. In essence, the collision operator is evaluated by linearizing about an appropriate equilibrium distribution, constructed so that

$$\rho = \sum_i n_i^{eq}, \quad \mathbf{j} = \sum_i n_i^{eq} \mathbf{c}_i, \quad \mathbf{\Pi}^{eq} = \sum_i n_i^{eq} \mathbf{c}_i \mathbf{c}_i = p\mathbf{1} + \rho\mathbf{u}\mathbf{u}. \tag{2.3}$$

The equation of state is $p = \rho c_s^2$, where $c_s = \sqrt{1/2}$ is the speed of sound.

Molecular collisions conserve mass and momentum but change the nonequilibrium (or viscous) part of the momentum flux $\mathbf{\Pi}^{neq} = \mathbf{\Pi} - \mathbf{\Pi}^{eq}$. In our simulations only the shear modes survive the collision process so that the post-collision momentum flux $\mathbf{\Pi}'$ is given by

$$\mathbf{\Pi}' = \mathbf{\Pi}^{eq} + (1 + \lambda)(\overline{\mathbf{\Pi}} - \overline{\mathbf{\Pi}}^{eq}), \tag{2.4}$$

where the overbar indicates the traceless part of the tensor. The parameter $\lambda$ controls the rate of relaxation of the stress tensor; it is related to the shear viscosity of the fluid,

$$\eta = -\rho(2/\lambda + 1)/6, \tag{2.5}$$

and lies in the range $-2 < \lambda < 0$. In the low-Reynolds-number limit, fluid inertia is neglected; in this case Eq. 2.4 can be simplified to

$$\mathbf{\Pi}' = p\mathbf{1} + (1 + \lambda)\overline{\mathbf{\Pi}}. \tag{2.6}$$

The post-collision distribution is given in terms of the mass density, momentum density, and updated momentum flux $\mathbf{\Pi}'$,

$$n_i + \Delta_i(n) = a_0^{c_i}\rho + a_1^{c_i} j_\alpha c_{i\alpha} + a_2^{c_i} \Pi'_{\alpha\beta}\overline{c_{i\alpha}c_{i\beta}} + a_3^{c_i}(\Pi'_{\alpha\alpha} - 3\rho c_s^2) : \tag{2.7}$$

the coefficients $a_i^{c_i}$ are model dependent; values for the 18 velocity model used in this work are given in Eq. 2.11 of Part I.

The solid particles are defined by a boundary surface, of any size or shape, which cuts some of the links between lattice nodes (see Fig.2 of paper I). The fluid particles moving along these links interact with the solid surface at boundary nodes placed halfway along the links. At each boundary node there are two incoming distributions $n_i(\mathbf{r}, t_+)$ and $n_{i'}(\mathbf{r} + \mathbf{c}_i, t_+)$, corresponding to velocities $\mathbf{c}_i$ and $\mathbf{c}_{i'}$ ($\mathbf{c}_{i'} = -\mathbf{c}_i$) parallel to the link connecting $\mathbf{r}$ and $\mathbf{r} + \mathbf{c}_i$; the notation $n_i(\mathbf{r}, t_+) = n_i(\mathbf{r}, t) + \Delta_i(\mathbf{r}, t)$ is used to indicate the post-collision distribution (Eq. 2.7). The velocity of the boundary node $\mathbf{u}_b$ is determined by the solid particle velocity $\mathbf{U}$, angular velocity $\mathbf{\Omega}$, and centre of mass $\mathbf{R}$,

$$\mathbf{u}_b = \mathbf{U} + \mathbf{\Omega} \times (\mathbf{r} + \tfrac{1}{2}\mathbf{c}_i - \mathbf{R}). \tag{2.8}$$

By exchanging population density between $n_i$ and $n_{i'}$ we can modify the local momentum density to match the velocity of the solid particle surface at the boundary node, without affecting either the mass density or the stress, which depend only on the sum $n_i + n_{i'}$. The precise form for the boundary-node collision operator is

$$n_i(\mathbf{r} + \mathbf{c}_i, t+1) = n_{i'}(\mathbf{r} + \mathbf{c}_i, t_+) + 2a_1^{c_i}\rho\mathbf{u}_b \cdot \mathbf{c}_i,$$
$$n_{i'}(\mathbf{r}, t+1) = n_i(\mathbf{r}, t_+) - 2a_1^{c_i}\rho\mathbf{u}_b \cdot \mathbf{c}_i; \tag{2.9}$$

As a result of the boundary-node interactions (Eq. 2.9), forces are exerted on the solid particles,

$$\mathbf{f}(\mathbf{r} + \tfrac{1}{2}\mathbf{c}_i, t+\tfrac{1}{2}) = 2[n_i(\mathbf{r}, t_+) - n_{i'}(\mathbf{r} + \mathbf{c}_i, t_+) - 2a_1^{c_i}\rho\mathbf{u}_b \cdot \mathbf{c}_i]\mathbf{c}_i; \tag{2.10}$$

thus momentum is exchanged locally between the fluid and the solid particle, but the combined momentum of solid and fluid is conserved. The forces and torques on the solid particle, obtained by summing $\mathbf{f}(\mathbf{r} + \tfrac{1}{2}\mathbf{c}_i)$ and $(\mathbf{r} + \tfrac{1}{2}\mathbf{c}_i) \times \mathbf{f}(\mathbf{r} + \tfrac{1}{2}\mathbf{c}_i)$ over all the boundary nodes associated with a particular particle, are then used to update the particle velocity and angular velocity, according to the laws of Newtonian mechanics. The mass and moment of inertia of the particle are preassigned, depending on the assumed distribution of mass within the particle. Since the forces at the boundary nodes are generated at the half-integer time steps, the force $\bar{\mathbf{f}}$ at the intermediate integer time,

$$\bar{\mathbf{f}}(\mathbf{r} + \tfrac{1}{2}\mathbf{c}_i, t) = (1/2)[\mathbf{f}(\mathbf{r} + \tfrac{1}{2}\mathbf{c}_i, t - \tfrac{1}{2}) + \mathbf{f}(\mathbf{r} + \tfrac{1}{2}\mathbf{c}_i, t + \tfrac{1}{2})], \tag{2.11}$$

is used to update the particle velocities every two timesteps:

$$\mathbf{U}(t+1) = \mathbf{U}(t-1) + 2M^{-1}\mathbf{F}(t), \quad \mathbf{\Omega}(t+1) = \mathbf{\Omega}(t-1) + 2\mathbf{I}^{-1} \cdot \mathbf{T}(t). \tag{2.12}$$

The particle mass $M$ and moment of inertia $\mathbf{I}$ are preassigned parameters which control the rate at which particles respond to the fluid flow; usually $M$ and $\mathbf{I}$ are on the order of several thousand (in lattice units). Since the velocities vary slowly on the time scale of a lattice-Boltzmann cycle, the precise form for the velocity update is usually not too important; however, it is important to use time-smoothed forces and torques, as described in Eq. 3.18 of Part I.

In this section we have briefly summarized the basic implementation of the simulation method. A detailed discussion of the method and its underlying theoretical justification is given in Part I. Methodology associated with specific applications will be described in the appropriate sections, along with the numerical results.

## III. LOW REYNOLDS NUMBER HYDRODYNAMICS

Low-Reynolds number hydrodynamics is applicable to an intermediate range of particle sizes. For instance, in a suspension of particles sedimenting in water under the influence of gravity, the appropriate size range is between approximately $1\mu$m and 1mm. If the particles are smaller than about $1\mu$m, Brownian motion is important and the effect of fluctuations in the fluid must be considered (see section VI); for particles larger than about 1mm, fluid inertia begins to have an effect (see section IV). In this section three different sets of problems are examined; periodic arrays of spheres, hydrodynamic interactions between two spheres, and the properties of random dispersions of spheres as measured by the short-time transport coefficients. The simulations were run under conditions such that the inertial contributions to the momentum flux were completely ignored (see Eq. 2.6). In all cases, results are compared with accurate numerical solutions of the Stokes equations,

$$\boldsymbol{\nabla} \cdot \mathbf{u} = 0, \quad \boldsymbol{\nabla} p = \eta \nabla^2 \mathbf{u}, \tag{3.1}$$

determined by multipole moment expansions of the force density on the particle surfaces (Ladd, 1988; Ladd, 1990).

### A. Periodic arrays

In this section we focus on the translational and rotational friction coefficients of a simple cubic lattice of spheres. The force on a steadily moving sphere is computed under conditions of vanishing volumetric flow rate (Batchelor, 1972), which is enforced in two different ways. In the first case a quasi-periodic system is used, as illustrated in Fig. 1. Each solid particle is given the same translational velocity, perpendicular to the boundary walls. Since the system is Galilean invariant, the measured drag force depends only on the velocity difference between the solid particles and the boundary walls. The most convenient choice is to set the boundary wall velocities to zero, so that the volumetric flow rate is zero. Then the translational friction coefficient is just the drag force on the central sphere divided by its velocity. It has been verified that these simulations are unconditionally stable for any particle velocity or flow rate (set by the boundary wall velocity), and that the friction coefficient is independent of velocity. For more than three cells, the drag force on the central sphere is independent of the number of cells in the quasi-periodic array, implying that three cells is enough to simulate a fully periodic system.

An alternate method is to hold the particles stationary (by considering them to be infinitely massive) and applying a uniform force density throughout the fluid, to simulate a pressure gradient across the system. For example, if a constant increment $\Delta j_x$ is applied to the $x$-component of momentum at each node, the macroscopic pressure gradient is given by $\nabla_x p = \Delta j_x$. The fluid velocity at each node is measured after half the force has been applied; it was found that this symmetrical procedure led to an identical result to that obtained with quasi-periodic systems. The measured drag force on the particle $F_D$ is smoothed over two successive time steps (see Eq. 2.11),

$$F_D = \sum_{\mathbf{r}_b} (1/2)[f_x(\mathbf{r}_b, t - \tfrac{1}{2}) + f_x(\mathbf{r}_b, t + \tfrac{1}{2})]; \qquad (3.2)$$

here $\mathbf{r}_b$ denotes the location of a boundary node. The balance of forces at steady state (for instance during sedimentation) implies that the total (gravitational) force on the sphere is balanced by the sum of the drag force and the buoyancy force $F_B = -V_S \Delta j_x$ ($V_S$ is the volume of the sphere); thus

$$F_D + F_B = (V - V_S)\Delta j_x, \qquad (3.3)$$

or equivalently $F_D = V \nabla_x p$. At steady state, the simulations satisfy this identity precisely. The volumetric flow rate $U_V$ is measured from the time-smoothed flow field (see Eq. 3.20 of paper I),

$$U_V = V^{-1} \sum_{\mathbf{r}_b} (1/4)[u_x(\mathbf{r}_b, t-1) + 2u_x(\mathbf{r}_b, t) + u_x(\mathbf{r}_b, t+1)]. \qquad (3.4)$$

Results from simulations at low solids volume fraction (less than 10%) are shown in Table II; numerical values from both quasi-periodic and pressure driven simulations were indistinguishable.

Since the particle surfaces are discrete, it is not possible to determine an accurate value of the particle radius *a priori*. An effective hydrodynamic radius $a$ can be computed from the drag coefficient $\xi^T = F_D/U_V$ of a simple cubic lattice of spheres at low solids concentration, using the asymptotic expression for the drag coefficient (Hasimoto, 1959);

$$\frac{6\pi\eta}{\xi^T} = \frac{1}{a} - \frac{2.837}{L} + \frac{4.19}{L^3}a^2 - \frac{27.4}{L^6}a^5. \qquad (3.5)$$

The first term on the right hand side is the Stokes result for an isolated sphere; the next three terms are the periodic corrections to the Stokes drag for a unit cell of length $L$. In Table II three different estimates of the radius of the solid particle are given. The first is the input radius $a_0$ which defines the boundary surface of the sphere; all lattice nodes inside this surface comprise the solid particle. Obviously, there is a range of values of $a_0$ which lead to the same object, for instance $\sqrt{2} < a_0 < \sqrt{3}$. Thus a reasonable and well defined measure of the size of the object is the radius of the equal volume sphere $a_V$. The effective radius, determined from Eq. 3.5, is also shown. All results in Table II are for $\lambda = -1$ (Eq. 2.4), corresponding to a kinematic viscosity $\nu = 1/6$ (Eq. 2.5). It can be seen that a consistent hydrodynamic radius emerges from these simulations, confirming that the simulated drag coefficients for dilute periodic suspensions are in agreement with Eq. 3.5. It should be emphasized that the same values for the effective radii of the objects are used in *all* subsequent simulations; the hydrodynamic radius $a$ is *not* treated as a variable parameter, although even better agreement between theory and simulation could be achieved by very small variations in $a$. However, the effective radius is not completely independent of the fluid viscosity. Over a useful range of viscosities ($10^{-3} < \nu < 1$), $a$ varies by about 1 lattice spacing (Table III). For large enough objects, this variation could be ignored, but in most practical instances, the effective radius must be calibrated for each fluid viscosity.

In order to test the simulation method at higher solids concentrations, the translational and rotational friction coefficients of a simple cubic lattice have been calculated over the entire density range. The rotational friction coefficient was determined by measuring the steady-state torque for a fixed rotational velocity. The results are shown for various size spheres in Fig. 2. The overall agreement is quite good; spheres of radius 5-10 lattice spacings give quantitative results over the whole concentration range. It is also possible to "tune" the particle shape to a certain extent. For instance, the rotational friction coefficient for a sphere of radius $a = 8.47$, corresponding to an input radius $a_0 = 8.5$ is about 10% too large at high solids concentrations . However, for a slightly smaller sphere $a = 8.44$ ($a_0 = 8.4$), the rotational friction is only about 5% too large. This difference arises because of the discrete lattice, which, in certain directions, blunts the surface of the sphere into flat plates which have different areas for the two different radius spheres. This difference in area has a much greater effect on the rotational friction than the translational friction, because the faces are in relative motion. In random dispersions, these differences are less severe than for the lattice arrangements, and no attempt has been made in this work to optimize the particle shape.

## B. Hydrodynamic interactions

A more stringent test of the simulations occurs if there is relative motion between the solid particles. In such cases there are singular lubrication forces when the particles are close to contact; forces along the line of centers diverge as $s^{-1}$ and forces perpendicular to the line of centers diverge as $\ln s^{-1}$, where $s = (R_{12} - 2a)/a$ is the gap between the particles relative to the radius. The simulations comprise a periodic unit cell, $2L \times L \times L$, with spheres located at $(L \pm L/2, L/2, L/2)$. The two spheres move with opposite velocities $\mathbf{u}$ and $-\mathbf{u}$; thus there is no net momentum flux. For all velocities, the drag force was found to be a linear and superposable function of $\mathbf{u}$. In Fig. 3 the simulation results are compared with integral-equation solutions (as in Ladd, 1988) for an identical geometry, including an explicit and exact calculation of the lubrication forces (Bossis and Brady, 1987). Again the overall agreement is good; the simulations are accurate for particle separations of 1 lattice spacing and beyond. However, the divergence of the lubrication forces near contact is not reproduced; instead the friction coefficients asymptote to a value more or less the same as that found at separations of 1 lattice spacing. These results could be improved upon by taking explicit account of the lubrication forces, as in Stokesian dynamics and related methods (Bossis and Brady, 1987; Ladd, 1990); however lattice-Boltzmann simulations of the short-range friction coefficients are much more accurate than typical multipole-moment or boundary integral methods and corrections for lubrication are not really necessary. Bulk properties of a suspension of such particles, which sample all distances, will be less sensitive to the short-range interactions than the single-configuration results reported here. It will be seen that accurate simulations of the transport properties of dense suspensions can achieved with quite small particles.

### C. Transport coefficients

In this section hydrodynamic transport coefficients of equilibrium distributions of spheres are reported. Experimentally these results correspond to short times, before any significant changes take place in particle configuration as a result of the externally imposed flow. We have computed the permeability of fixed random arrays of spheres ($K$), the collective mobility ($\mu$) or sedimentation velocity, the short-time self diffusion coefficient ($D_s$), and the high-frequency viscosity ($\eta_\infty$). In Fig. 4, results from lattice-Boltzmann simulations are compared with independent simulations based on multipole moment expansions of the Oseen equation (Ladd, 1990); these latter results, when corrected for finite size effects, are in excellent agreement with experiment. Here we compare results for small periodic systems containing 16 spheres per unit cell in all cases; thus the plots in Fig. 4 differ somewhat from those reported earlier (Ladd, 1990), because of finite size effects. Typically, we average over 10–100 different configurations of spheres for each calculation; the individual configurations were generated with a standard hard-sphere Monte Carlo program (Allen and Tildesley, 1987). Thus we have a relatively rapid method of checking the accuracy of the simulations as a function of volume fraction $\phi$ and particle size $a$. The results are clustered around three typical concentrations; dilute ($\phi \simeq 0.05$), semi-dilute ($\phi \simeq 0.25$), and concentrated ($\phi \simeq 0.45$).

The permeability of a fixed array of spheres relates the volume-averaged velocity ($\mathbf{U}_V$) of the fluid (Eq. 3.4) to the pressure gradient

$$\mathbf{U}_V = -\frac{K}{\eta}\boldsymbol{\nabla} p; \tag{3.6}$$

the velocity field is averaged over the whole volume of the system, including the interior of the spheres where the fluid is at rest. Thus the permeability of a random array of spheres is calculated in a similar fashion to the translational friction coefficient of an isolated sphere; each sphere in the sample is held fixed, and a uniform force density $\nabla p = \Delta \mathbf{j}$ is applied to every node at every time step. Numerical results show that the lattice-Boltzmann simulations predict the permeability of random dispersions of spheres rather well. Particles with radii as small as 2.5 lattice spacings are just as accurate as Stokesian dynamics (Phillips et al., 1988), although larger spheres ($a > 5$) are needed for accurate results at high concentrations.

The collective mobility (or sedimentation velocity) and self-diffusion coefficient also converge rapidly with particle radius, as shown in Fig. 4. It can be seen that the lubrication forces, which contribute in an average way to the self-diffusion coefficient, are quantitatively accounted for by the simulations. The methodology for determining the collective and self-mobilities is as follows. An external force $F_{ext}$ is applied to one particle (for $D_s = k_B T \mu_s$) or is divided equally among all $N$ particles (for $\mu$); this external force is balanced by a force density $\Delta j = -F_{ext}/V$ which is applied to all nodes in the fluid: thus

$$\frac{6\pi\eta \langle U_1 \rangle}{F_{ext}} = \mu_s, \quad \frac{6\pi\eta \left\langle \sum_{i=1}^{N} U_i \right\rangle}{F_{ext}} = \mu. \tag{3.7}$$

Many of the results for $D_s$ were computed by an alternate and comparable method in which a force $F_{ext}$ is applied to one sphere and a balancing force $-F_{ext}/(N-1)$ is applied to all the other spheres. By measuring the average particle velocity of the tagged sphere $<U_1>$, we then obtain $\mu_s$ from the relation

$$\frac{6\pi\eta <U_1>}{F_{ext}} = \mu_s - \frac{(\mu - \mu_s)}{N-1} = \frac{\mu_s N}{N-1} - \frac{\mu}{N-1}. \tag{3.8}$$

To simulate the motion of particles under the action of an external force, the fixed particle constraint must be removed and the particle velocities updated according to Eq. 2.12. In these calculations the mass and inertia were set to between 2 and 4 times that of an equal volume of fluid. The total force $\mathbf{F}$ is given by the sum of hydrodynamic force and the external force; at steady-state, these forces balance and the particle velocities are constant.

The code is unstable for small values of either the mass or the inertia, because the boundary conditions (Eq. 2.9) reflect most of the incoming momentum. From a molecular point of view this corresponds to the assumption that the solid particle is much more massive than the fluid molecules, which is not invariably so. It is straightforward to establish an approximate stability criterion. For a quiescent fluid (*i.e.* ignoring variations in $f_i$), the momentum transfer per unit area per time step is approximately $-2\rho \mathbf{u}_b$. Integrating over the surface of the sphere we find, from Eq. 2.12, that the change in particle velocity $\Delta \mathbf{U}$ is given by

$$\Delta \mathbf{U} \simeq -\frac{6}{\beta a}\mathbf{U}, \tag{3.9}$$

where $\beta = \rho_s/\rho$ is the ratio of solid particle mass density to fluid mass density. A similar expression holds for the angular velocity; the numerical factor in this case is 10. Thus we expect stable solutions if $\beta a > 10$; it has been verified empirically that this is indeed correct.

The high-frequency viscosity for fixed configurations of spheres in an external shear flow has also been computed. We have not, as yet, been able to devise a homogeneous shear algorithm, analogous to the those used for molecular liquids (Hoover et al., 1980; Ladd, 1984). Thus we again use a quasi-periodic system with a geometry similar to that shown in Fig. 1, but with many spheres in each unit cell. Of course the configuration of spheres in each cell is the same. A shear flow is set up by moving one of the walls at a fixed velocity, $u_y$, parallel to the wall. The stress and velocity profile in the central cell are measured, and from this the viscosity can be determined; once again results for quasi-periodic systems of 3 unit cells and 5 unit cells are essentially identical. The stress in the suspension includes contributions from the fluid stress modes, together with particle fluid interactions; these are computed in a similar way to the torques, by summing $\mathbf{r}_b\overline{\mathbf{f}}(\mathbf{r}_b)$ over the particle surface. As before the simulations manage to pick up the effects of lubrication quite well, although at the highest volume fraction the smaller sphere ($a \simeq 2.5$) is inadequate. A drawback of the present scheme for imposing an external shear flow is that, at higher solids concentrations, the velocity gradient is non-uniform in the outer two cells, although it is quite uniform in the central cell (Fig. 5). Thus the time it takes for the system to reach steady state can be quite long, as many as $10^5$–$10^6$ time steps, during which time a steady velocity profile is slowly realized. It is to be hoped that a better choice of initial conditions will speed up the approach to steady state considerably. However, the slow approach of the initial configuration to steady state will be less of a hindrance in dynamical simulations, which can use long time averages from a single initial configuration instead of ensemble averages over many initial configurations.

### D. Shared nodes

A technical difficulty arises in the simulation of many-particle suspensions which has not yet been discussed. As shown in Fig. 2 of paper I, boundary nodes sit at the midway points between lattice nodes. Thus some fraction of the boundary nodes can be shared between two different particles, with no intervening fluid between them. The number of "shared nodes" is generally small compared with the total number of boundary nodes; at fixed volume fraction the ratio is proportional to $a^{-3}$. At the highest concentrations (45%) the proportions of shared nodes are 50%, 5%, and 1% for spheres of radius 2.5, 4.5, and 8.5 respectively. At lower concentrations, the proportion of shared nodes declines rapidly; thus for radius 2.5 spheres the proportion of shared nodes at solids concentrations of 45%, 25%, and 5% is 50%, 8%, and less than 1% respectively. Although the specific details of the shared nodes are unimportant for sufficiently large particles ($a > 10$), the convergence is markedly improved by a careful treatment of these situations.

To handle the shared nodes correctly, the effects of the local velocity of both particles must be taken into account. Of course, these velocities are similar (at steady state) so the boundary node velocity can be taken to be the average of that computed from the velocities of each particle (*c.f.* Eq. 2.8)

$$\mathbf{u}_b = \frac{1}{2}\left[\mathbf{U}_i + \mathbf{\Omega}_i \times (\mathbf{r}_b - \mathbf{R}_i) + \mathbf{U}_j + \mathbf{\Omega}_j \times (\mathbf{r}_b - \mathbf{R}_j)\right]. \tag{3.10}$$

Using this velocity, the fluid populations are updated (Eq. 2.9) and the force is computed (Eq. 2.10). This force is then divided evenly between the two particles. The numerical results for small particles ($a < 5$) are quite sensitive to the exact prescription for dividing up the force; variations on the even division of force (for instance assigning the force first to one sphere and then, on the next time step, to the other) can produce significantly different transport coefficients for small spheres at high solids concentrations.

## IV. FINITE REYNOLDS NUMBER FLOWS

The simulation technique can be carried over, without modification, to flows at non-zero Reynolds numbers. However, the simplified form for the equilibrium distribution (Eq. 2.39 of paper I) cannot be used; the appropriate distribution is given in Eq. 2.5 of paper I. Two steady-state flows, up to Reynolds number $R_e = 200$ have been studied; flows past a column of cylinders and flows past a simple-cubic array of spheres. Our results are compared with finite difference and finite element solutions of the Navier-Stokes equations.

### A. Column of cylinders

The problem of fluid flow past a circular cylinder has been studied by many authors. Since our simulation method is designed for fairly dense suspensions, its use of a regular mesh of lattice nodes makes it rather inefficient for studying flows past isolated objects. A reasonable compromise is a column of closely spaced cylinders (Fornberg, 1991), which can be simulated by a long narrow channel with periodic boundary conditions. Our simulations are restricted to a single geometry, namely a column of cylinders separated by 10 cylinder radii. Test calculations with a small cylinder ($a = 2.1$) showed that drag coefficients within 1% of the infinite channel length result could be obtained for a channel length $L = 16W = 160a$. The flow was driven by setting the velocity of a single row of nodes to the desired input velocity $U_0$; the cylinder was placed $4W = 40a$ downstream, leaving a maximum wake length of $12W = 120a$. Simulations were run with cylinders of input radii $a_0 = 1.5$, $a_0 = 4.5$, and $a_0 = 9.5$ with a kinematic viscosity $\nu = 0.01$. The effective hydrodynamic radii of the cylinders were determined to be 2.1, 5.1, and 10.1 respectively. The calibration of the cylinder size was performed in an analogous way to that for spheres, by measuring the drag force of a dilute periodic array of cylinders at $R_e = 0$. To a first approximation (in solids concentration) the friction coefficient $\xi^T$ is given by the two-dimensional equivalent of Eq. 3.5

$$\frac{2\eta}{\xi^T} = \frac{1}{A} \sum_{\mathbf{k} \neq 0} \left[\frac{J_0(ka)}{k}\right]^2 ; \qquad (4.1)$$

$A$ is the area of the unit cell, $\mathbf{k} = 2\mathbf{n}\pi/L$ is the wavevector, and $J_0$ is a cylindrical Bessel function.

A contour plot of the stream lines for the larger ($a = 10.1$) cylinder is shown in Fig. 6 at various Reynolds numbers ($R_e = 2U_0 a/\nu$); the maximum Mach number (at $R_e = 100$) is 0.07. The flow at $R_e = 100$ is eventually unstable; however, the instability is sufficiently delayed that a quasi-steady state could be reached before its onset. A detailed comparison of the simulation results with Fornberg's finite difference code is shown in Table IV.[1] It can be seen that the lattice-Boltzmann simulations reproduce all the measured characteristics of the flow quantitatively, including the minimum in the stream function at the center of the recirculation zone.

### B. Simple-cubic lattice of spheres

These simulations are similar to the calculations of the creeping-flow translational friction coefficient described in section IIIA. The sample contains a single stationary sphere of radius $L/2$ in a cubic box of length $L$ with periodic boundary conditions; the flow is driven by a uniform force density, as before. The only difference is that the full non-linear form for the equilibrium distribution is used, with a kinematic viscosity $\nu = 0.01$; the Mach number is then small, typically of the order of $10^{-2}$. The results are reported in Table V. The input radii ($a_0 = L/2 - 0.2$) were determined so that the effective radius $a$ was as close as possible to $L/2$; the choice of input radius is consistent with the result reported in Table III for $\nu = 1/96$.

---

[1] Fornberg's published results (Fornberg, 1991) begin at $R_e = 100$. Because his code utilizes the symmetry present in steady flows, he obtains time independent solutions at all Reynolds numbers. Our simulations exhibit vortex shedding at higher Reynolds numbers; we can only obtain steady solutions up to $R_e = 100$. Dr. Ramesh Natarajan kindly ran some intermediate Reynolds numbers for these comparisons.

It can be seen that the mean flow velocity $U_V$ (or Reynolds number) for a fixed drag force ($F_D = -V\nabla p$) converges quite rapidly with increasing particle radius (or unit cell length $L$). Thus it is surprising that the simulation results do not agree at all with the results of Lahbabi and Chang, 1985, who studied the identical problem via a truncated mode expansion of the Navier-Stokes equations. A comparison is shown in Fig. 7; the reduced pressure gradient $\nabla p a^3/\rho\nu^2$ is plotted vs Reynolds number $R_e = 2U_V a/\epsilon\nu$, where $\epsilon = 1 - \pi/6$ is the porosity[2]. At Reynolds numbers greater than about 40 the simulations deviate rapidly from Lahbabi and Chang's results; they differ by about a factor of two at $R_e = 200$, with the simulations predicting a much smaller pressure drop. To investigate this discrepancy further, a finite element model of the problem (Tompson, 1983) has been run; these results, shown as open symbols in Fig. 7, are in excellent agreement with the lattice-Boltzmann simulations[3]. The lattice-Boltzmann/finite element results together suggest that the calculation of Lahbabi and Chang is incorrect. Moreover, they indicate that finite Reynolds number corrections to the Stokes drag in a regular arrays of spheres are considerably smaller than in random arrays; only about 20% at $R_e = 100$.

---

[2] Note that Lahbabi and Chang use a different definition of $R_e$, based on the particle radius rather than the diameter; thus their Reynolds number is smaller than ours by a factor of 2.

[3] The finite element code LAMFLOW used for these comparisons is based on a penalty-function formulation of the time-independent Navier-Stokes equations (Hughes et al., 1979). The calculations used 148, 8-node prismatic elements to represent one quadrant of the periodic unit cell. The code was recently resurrected for this problem by its author, Dr. Andy Tompson

## V. TIME-DEPENDENT HYDRODYNAMIC INTERACTIONS

The classical description of colloidal particle dynamics is the Einstein-Smoluchowski equation, in which the details of the short time dynamics are ignored. The hydrodynamic interactions are assumed to be fully developed; in other words there is a complete separation of time scales between the dynamics of the fluid and the (diffusive) motion of the solid particles. In reality, hydrodynamic interactions develop by the diffusion of fluid vorticity, with a characteristic time scale $\tau_f = a^2/\nu$. Lattice-gas/lattice-Boltzmann simulations methods are motivated by the observation that, by solving time-dependent Navier-Stokes equations, the correct hydrodynamic forces and torques can be computed from purely local interactions; thus the computational cost scales linearly with the system size. However, for low Reynolds number flows, an extra time scale separation between the solid and fluid motions must now be introduced, to allow time for the steady-state hydrodynamic interactions to develop. In practice this is not a severe constraint, as can be seen from the following considerations. The time scale characterizing the particle motion is $\tau_s = a/U$, where $U$ is a typical particle velocity; the ratio $\tau_f/\tau_s$ is just the Reynolds number $R_e$. It is known that the creeping-flow limit is generally equivalent to conditions in which $R_e < 1$ in dense suspensions and $R_e < 0.1$ in dilute suspensions (Happel and Brenner, 1986); for a particle radius $a = 5$ and typical kinematic viscosity $\nu = 1/6$, this implies particle velocities less than 0.01. Thus a typical dynamical simulation (Ladd, 1993a), where the particles must travel distances on the order of 100 radii, would require about $10^5$ time steps. Based on a particle radius $a = 4.5$, a complete sedimentation run would require an estimated total of $10^8$ node updates per particle; a modern workstation (IBM 580 or HP 735) can update more than $10^5$ nodes per second, so that a complete run would take about 15 minutes per particle. Moreover, the code can be readily executed, in parallel, on many processors at once. Thus simulations of $10^5$ or more spheres should be feasible on a massively-parallel supercomputer. Even with the present code, systems of 128 spheres (see section VI) are simulated routinely (mainly on SPARC I and SPARC II computers), and in a few instances systems of 1024 spheres (on an IBM 580).

Although the original motivation for developing a method that simulated time-dependent hydrodynamic interactions was purely computational, the emergence of diffusing-wave spectroscopy (Weitz et al., 1989) has initiated experimental studies into the motion of colloidal particles at very short times. Underlying the experimentally observed behavior (Zhu et al., 1992; Kao et al., 1993) are both time and space dependent hydrodynamic interactions; with this new simulation technique we have a unique capability to determine these interactions *ab initio* (Ladd, 1993c). Since these experiments track the Brownian motion of the colloidal particles, a brief discussion of the relevant simulations will be deferred to section VI; an account of this work has already been reported (Ladd, 1993c). Here we focus on studies of the motion of isolated spheres, to verify the accuracy of the simulations in tracking the temporal and spatial development of the hydrodynamic forces (see also Fig. 5 of paper I).

The unsteady motion of an isolated sphere has been worked out in some detail; the basic results for low Reynolds number (ignoring fluid inertia) can be found, for example, in Landau and Lifshitz, 1959. Recently this work has been extended to small but finite Reynolds number (Lovalenti et al., 1993); however, at present our results are limited to the time dependent $R_e = 0$ case only. We begin by investigating the motion of a sphere undergoing small amplitude rotational oscillations $\Omega(t) = \Omega_0 \cos(\omega t)$; the in phase and out of phase components of the drag torque were measured to determine the modulus of the torque $|T|$ and the phase lag $\psi$ (see Fig. 8). Since there is fluid both inside and outside the sphere, contributions to the drag torque from both regions are summed together to give the solid lines in Fig. 8. These expressions, in terms of the reduced frequency $\omega^* = \omega a^2/\nu$ and the zero frequency drag coefficient $\xi^R = 8\pi\eta a^3$, are (Landau and Lifshitz, 1959)

$$T_{ext} = -\xi^R \Omega_0 \left\{ 1 - \frac{\imath\omega^*}{3[1 + \sqrt{-\imath\omega^*}]} \right\},$$

$$T_{int} = \xi^R \Omega_0 \left\{ 1 - \frac{\imath\omega^*}{3[1 - \sqrt{\imath\omega^*}\cot(\sqrt{\imath\omega^*})]} \right\}. \quad (5.1)$$

It can be seen (Fig. 8) that the simulated phase lag is inaccurate at sufficiently high frequencies, when the period of oscillation $(2\pi/\omega)$ is less than about 10 simulation time steps. High frequency motion can be accurately simulated either by increasing the size of the spheres or by decreasing the viscosity of the fluid. However, reduced frequencies ($\omega a^2/\nu$) larger than 10 are unimportant in most instancies.

At low frequencies, the drag forces and torques from the interior fluid reduce to the additional inertial drag of the oscillating mass of fluid; for rotational oscillations the interior torque becomes $T_{int} = 8\pi\rho a^5 \imath\omega\Omega^0/15$ (Eq. 5.1). Since the added inertia from the interior fluid can be readily accounted for (as will be seen shortly), only the deviations in the interior drag force from the inertial drag are significant; these deviations are proportional to $\omega^2$, to leading order. Thus the effects of the interior fluid can be minimized by using a smaller viscosity in the exterior fluid than in the interior fluid as shown in Fig. 9. Here the interior fluid viscosity is kept fixed at $\nu = 1/6$ whereas the exterior fluid viscosity is varied between 1/24 and 1/96. It can be seen that for the same size sphere ($a_0 = 4.5$) agreement between simulation and theory improves rapidly with increasing viscosity ratio (see also results in Fig. 8 for $a = 4.53$); moreover deviations of the drag torque from the asymptotic form (exterior torque plus interior inertia) become small. However, it will be seen that this refinement is unnecessary in time-domain simulations, since these high frequencies play a very small role.

In Fig. 10 the decay of an initial translational velocity $U(0)$ or rotational velocity $\Omega(0)$ of a sphere is compared with theoretical predictions based on an inverse Laplace transform of the frequency-dependent equations of motion (Hauge and Martin-Löf, 1973)

$$U(\omega) = \frac{U(0)}{-\imath\omega(M + M_f) + \xi^T(\omega)},$$
$$\Omega(\omega) = \frac{\Omega(0)}{-\imath\omega(I + I_f) + \xi^R(\omega)}. \quad (5.2)$$

The friction coefficients that appear in Eqs. 5.2 are the exterior friction coefficients

$$\xi^T(\omega) = \xi^T\left\{1 + \sqrt{-\imath\omega^*} - \imath\omega^*/9\right\}$$
$$\xi^R(\omega) = \xi^R\left\{1 - \frac{\imath\omega^*}{3[1 + \sqrt{-\imath\omega^*}]}\right\}; \quad (5.3)$$

the internal drag forces are approximated as purely inertial mass; $\imath\omega M_f = 4\pi\rho a^3 \imath\omega/3$ for the translational friction and $\imath\omega I_f = 8\pi\rho a^5 \imath\omega/15$ for the rotational friction.

The decay of the translational velocity of a sphere can be expressed analytically (for example Hinch, 1975), but the rotational velocity has not been computed over the full time domain. Thus we follow Hauge and Martin-Löf, 1973 and convert the inverse Laplace transform to a definite integral (by contour integration) and then evaluate this integral by numerical quadrature to get the solid lines in Fig. 10. The mass and inertia of the solid particle comprise the sum of the assigned mass $M$ or inertia $I$, used to update the particle velocities (Eq. 2.12), and the mass $M_f = 4\pi\rho a^3/3$ or inertia $I_f = 8\pi\rho a^5/15$ of the interior fluid. Although there are no free parameters in the comparison of simulation with theory, the agreement is essentially perfect over the whole time domain. This supports the earlier assertion that the interior fluid contributes an inertial force, due to its extra mass, and very little else.

It is appropriate at this point to comment on the normalization of the theoretical results; they are not normalized to the $t = 0$ velocities $U(0)$ and $\Omega(0)$ but to something slightly less. There are two reasons for this. First, since the initial velocity is applied only to the particle and not to the interior fluid also, translational momentum $M_f U(0)$ and rotational momentum $I_f \Omega(0)$ are quickly lost to the internal fluid which thereafter moves essentially as a rigid body. Second, the theoretical results apply to an incompressible fluid; thus there is the well known "added-mass" effect which accounts for the momentum carried off by sound waves in a real fluid. Our lattice-Boltzmann simulations are slightly compressible; thus a fraction of the initial momentum $(M_f U(0)/2)$ is carried off almost instantaneously, after which the velocity tracks the renormalized incompressible theory. Of course, there is no added mass for the rotational velocity. Thus the overall normalizations at $t = 0$ are $M/(M + 3M_f/2)U(0)$ and $I/(I + I_f)\Omega(0)$.

## VI. FLUCTUATIONS

Suspensions of sub-micron sized particles undergo Brownian motion, due to the thermal fluctuations in the fluid. It has been shown (Ladd, 1993c), and further discussed in paper

I, that fluctuations can be incorporated into lattice-Boltzmann simulations by the straightforward addition of a random component to the fluid stress tensor. Thus the post-collision momentum flux in Eq 2.4 has in addition a fluctuating stress, $\boldsymbol{\sigma}'$, sampled from a Gaussian distribution and uncorrelated in space and time (Landau and Lifshitz, 1959),

$$\left\langle \sigma'_{\alpha\beta}(\mathbf{r},t)\sigma'_{\gamma\delta}(\mathbf{r}',t') \right\rangle = A\delta_{\mathbf{rr}'}\delta_{tt'}\left(\delta_{\alpha\gamma}\delta_{\beta\delta} + \delta_{\alpha\delta}\delta_{\beta\gamma} - (2/3)\delta_{\alpha\beta}\delta_{\gamma\delta}\right); \qquad (6.1)$$

the variance $A$ is proportional to the temperature of the fluctuating fluid. The exact relationship between the coefficient $A$ and the effective temperature is discussed in paper I (Eq. 4.18).

As a first test of the fluctuating lattice-Boltzmann simulations, the motion of an isolated sphere has been studied, at sufficiently short times that the periodic boundary conditions have no noticeable effect on the dynamics. The velocity correlation function of a spherical particle decays algebraically, with an asymptotic $t^{-3/2}$ dependence which arises from hydrodynamic correlations in the fluid (Alder and Wainwright, 1970); here the fluctuating lattice-Boltzmann equation is tested to make sure that it can account for the hydrodynamic memory effects that lead to long-time tails. In Fig. 11 it can be seen that, within the statistical error bars, the normalized velocity correlation functions are identical to the steady decay of the translational and rotational velocities of the sphere; thus our simulations satisfy the fluctuation-dissipation theorem. Once again there are no adjustable parameters in these comparisons.

Next we consider the motion of suspensions of colloidal particles at very short times, prior to the onset of Brownian motion. Diffusing-wave spectroscopy (Zhu et al., 1992) has shown that if the mean-square displacement is normalized by the self-diffusion coefficient times the time $\left\langle \Delta R^2(t) \right\rangle / 6D_s(\phi)t$, and plotted vs. a reduced time $t/\tau$, then, for all solids volume fractions, there is a scaling time $\tau(\phi)$ which collapses the experimental data onto one master curve, indistinguishable from the isolated-sphere result. Moreover $\tau$ seems related to the time it takes fluid vorticity to diffuse over a particle radius; values of $\tau$ determined from the scaled mean-square displacements are in good agreement with independent estimates of the vortex diffusion time $\eta_\infty(\phi)/\rho a^2$, based on the high-frequency suspension viscosity $\eta_\infty$. We have discussed this scaling in an earlier work (Ladd, 1993c); here we just point out that the simulations reproduce the experimentally observed scaling over the whole time and density range. Simulation data for a number of Brownian particles ($N = 128$) under the same scaling is shown in Fig. 12. Results with different size systems ($N = 16$ and $N = 1024$) indicate that the periodic boundary conditions have a negligible effect on the 128-sphere results for times up to about $100\tau$. The scaled data at various volume fractions collapse onto the dilute (single-particle) result, in excellent agreement with experiment both at very short times (Kao et al., 1993) and at somewhat longer times (Zhu et al., 1992). Moreover, the self-diffusion coefficient and viscosity that are required to scale the mean-square displacement are in quantitative agreement with independent simulations and experimental data (see Ladd, 1990). A comparison is shown in Fig. 13.

When the fluctuating lattice-Boltzmann equation is applied to a system of solid particles, it is found that the translational and rotational velocities of the particles come into approximate thermal equilibrium with the fluctuations in the fluid. Because our model does not conserve the total energy, but instead preserves a balance, on average, between viscous and fluctuating forces, there is no equipartition principle to force an even division of energy between all the modes in the system. Thus we do not find an exact thermal equilibrium between the characteristic temperature of the fluid and solid motions (see Table VI). In addition, the average kinetic energy of the solid particles is weakly dependent on the particle size and solids concentration; we find in general that the temperature characterizing translation and rotation are similar, and that they are typically 10–20% less than the effective temperature of the fluid fluctuations. For studies of particle diffusion, temperature will be taken to be defined by the mean translational kinetic energy; for the shear viscosity, there is some ambiguity in defining the temperature which will be discussed later.

The transport coefficients of a suspension of solid particles can be determined from the Green-Kubo relations (Hansen and McDonald, 1986). The simplest is the self-diffusion coefficient, which can be computed from the average mean-square displacement $\Delta \mathbf{R}$, or from the velocity autocorrelation function;

$$D_s = \lim_{t\to\infty} \frac{1}{6t} \langle \Delta \mathbf{R}(t) \cdot \Delta \mathbf{R}(t) \rangle = \frac{1}{3} \int_0^\infty \langle \mathbf{U}(t) \cdot \mathbf{U}(0) \rangle \, dt. \qquad (6.2)$$

For comparison purposes, it is simpler to look at normalized correlation functions

$$f_D(t) = \frac{\langle \mathbf{U}(t) \cdot \mathbf{U}(0) \rangle}{\langle \mathbf{U}(0) \cdot \mathbf{U}(0) \rangle}, \qquad (6.3)$$

where macroscopically $\langle \mathbf{U}(0) \cdot \mathbf{U}(0) \rangle = 3k_B T/M$; this expression is used to define the temperature of the suspension. A relaxation time $\tau_D$ can be defined as the integral of the normalized velocity correlation function

$$\tau_D = \int_0^\infty f_D(t); \qquad (6.4)$$

thus the ratio $D_s/D_0 = \tau_D/\tau_0$, with $\tau_0 = M/(6\pi\eta a)$. The integral in Eq. 6.4 and similar fluctuation expressions should be interpreted as a discrete quadrature; *i.e.*

$$\int_0^\infty f_D(t) dt \equiv f_D(0) + 2 \sum_{t=1}^\infty f_D(2t), \qquad (6.5)$$

where the factor of 2 arises because particle velocities are updated every two time steps.

The ratio $D_s/D_0$ is shown in Fig. 14 for various volume fractions and particle sizes, using systems of 16 and 128 spheres. The agreement is quite good, but, compared with the steady non-equilibrium flows (Fig. 4), larger particles are required at high densities for the same accuracy. This implies that fluctuation-dissipation is not exactly satisfied at high solids concentrations (otherwise both measures of the diffusion coefficient would be the same); I suspect these discrepancies come from the effects of the shared nodes (section IIID), which may well behave differently for fluctuating fluids than for steady flows. However, there is good agreement for the larger spheres, where the contribution from the shared nodes is negligible.

The collective or mutual diffusion coefficient $D$ is related to fluctuations in the total velocity of the solid phase

$$\mathbf{U}_c = \sum_{i=1}^N \mathbf{U}_i. \qquad (6.6)$$

The relationship between the fluctuations in $\mathbf{U}_c$ and the diffusion coefficient $D$ is discussed in the Appendix. Here we just quote the final result used for comparison with creeping-flow theory,

$$\mu/\mu_0 = \frac{1}{X_f \tau_0} \int_0^\infty \frac{\langle \mathbf{U}_c(t) \cdot \mathbf{U}_c(0) \rangle}{\langle \mathbf{U}_c(0) \cdot \mathbf{U}_c(0) \rangle} dt, \qquad (6.7)$$

where $X_f$ is the mass fraction of the fluid. The results (Fig. 14) are quite comparable to the earlier results for collective mobility (Fig. 4) even at high concentration. This is further evidence that the relatively slow convergence of the self-diffusion coefficients in Fig. 14 arises from the shared nodes, which they play a minor role in collective diffusion because of the absence of lubrication.

In section IV of paper I, it was shown that a discretized Green-Kubo relation could be derived relating the viscosity of the pure fluid to the equilibrium stress fluctuations; in our present notation, Eq. 4.13 of paper I can be written as

$$\eta V k_B T = \int_0^\infty \left\langle \Sigma^f{}_{xy}(t) \Sigma^f{}_{xy}(0) \right\rangle dt, \qquad (6.8)$$

where $\mathbf{\Sigma}^f$ is the fluid stress tensor. Equation 6.8 can be applied to solid-fluid suspensions by including both the fluid stress tensor (Eq. 4.12 of paper I) and the solid-fluid stress $\mathbf{\Sigma}^s$, summed over all the solid-particle surfaces. This latter contribution is computed in a similar way to the particle torques, summing the symmetric components of $\mathbf{r}_b \overline{\mathbf{f}}(\mathbf{r}_b)$ over all the boundary nodes. The fluid stress correlation function and the total stress ($\mathbf{\Sigma}^t = \mathbf{\Sigma}^f + \mathbf{\Sigma}^s$) correlation function have been computed; by integrating these correlation functions the viscosity ratio $\eta_\infty(\phi)/\eta$ can be computed (Eq. 6.8). However, the estimated high-frequency viscosities for the suspension are then consistently too low, by up to 20% at high volume fraction. The reason is that energy is not uniformly distributed between the particles and the fluid, so that the effective temperature in Eq. 6.8 is different for the particle contributions and the fluid contributions. We have corrected for this by using the fluid temperature for the fluid-fluid contribution and the particle temperature (see Table VI) for the particle-fluid terms. A comparison of viscosities is shown in Fig. 15. Here we have averaged over all 5 components of the shear stress; in this case these angle-averaged viscosities are very insensitive to the number of particles so that results for 16 particles and 128 particles are indistinguishable. The agreement is again good, although the method of computation is not totally satisfactory. It may transpire that viscosities can only be computed reliably by some kind of external flow simulation as in section IIIC, and thus we would have to reserve the Green-Kubo method for quantities, like diffusion, which involve only particle-particle correlation functions. The question of the thermal equilibrium between solid and fluid needs to be researched in more detail; hopefully a method of including fluctuations will be found which succeeds in bringing about a true equipartition of energy between solid and fluid.

## VII. CONCLUSIONS

The combination of molecular dynamics for the particulate phase and the fluctuating lattice-Boltzmann equation for the fluid phase has been shown to be a viable technique for quantitative simulations of hydrodynamically interacting particles. Even using small solid particles, with radii less than 5 lattice spacings, accurate results for hydrodynamic transport coefficients (permeability, sedimentation velocity, self-diffusion coefficient and viscosity) have been obtained over the whole range of packing fractions. The method is also very flexible; the particle size and shape, the electrostatic interactions, the flow geometry, the Peclet number and the Reynolds number, can all be varied independently. The technique is valid, without modification, at finite Reynolds number; lattice-Boltzmann simulations of flow past a column of cylinders are in quantitative agreement with finite-difference solutions at Reynolds numbers up to 100. The results for a dense array of spheres are in excellent agreement with finite-element calculations, but disagree with the truncated-mode analysis of Lahbabi and Chang. The simulations suggest that fluid inertia has a much smaller effect in periodic arrays than in random media.

It has been shown that fluctuations can be incorporated into the lattice-Boltzmann equation in a very straightforward fashion, and then used to simulate the dynamics of colloidal particles in suspension. The short-time transport coefficients of these suspensions can be accurately determined from autocorrelation functions of the fluctuations. We plan to examine long-time transport coefficients in the future, by tracking the motion of particles moving under the influence of Brownian forces.

## ACKNOWLEDGMENTS

I would like to acknowledge helpful conversations with Dr. Bengt Fornberg (Exxon Research, Annandale, New Jersey) and Dr. Peter Hoogerbrugge (Shell Research, Rijswijk, The Netherlands). I would also like to thank Dr. Ramesh Natarajan (IBM, Yorktown Heights, New York) and Dr. Andrew Tompson (Lawrence Livermore National Laboratory, Livermore, California) for unpublished results of flows past cylinders and spheres.

This work was supported by the U.S. Department of Energy and Lawrence Livermore National Laboratory under Contract No. W-7405-Eng-48.

TABLES

TABLE I. Labeling of the 18 velocity model. For each label $i$, the velocity vector and speed (in lattice units) are shown.

| $i$ | $c_x$ | $c_y$ | $c_z$ | $c$ |
|---|---|---|---|---|
| 1  |  1 |  0 |  0 | 1 |
| 2  | -1 |  0 |  0 | 1 |
| 3  |  0 |  1 |  0 | 1 |
| 4  |  0 | -1 |  0 | 1 |
| 5  |  0 |  0 |  1 | 1 |
| 6  |  0 |  0 | -1 | 1 |
| 7  |  1 |  1 |  0 | $\sqrt{2}$ |
| 8  | -1 | -1 |  0 | $\sqrt{2}$ |
| 9  |  1 | -1 |  0 | $\sqrt{2}$ |
| 10 | -1 |  1 |  0 | $\sqrt{2}$ |
| 11 |  1 |  0 |  1 | $\sqrt{2}$ |
| 12 | -1 |  0 | -1 | $\sqrt{2}$ |
| 13 |  1 |  0 | -1 | $\sqrt{2}$ |
| 14 | -1 |  0 |  1 | $\sqrt{2}$ |
| 15 |  0 |  1 |  1 | $\sqrt{2}$ |
| 16 |  0 | -1 | -1 | $\sqrt{2}$ |
| 17 |  0 |  1 | -1 | $\sqrt{2}$ |
| 18 |  0 | -1 |  1 | $\sqrt{2}$ |

TABLE II. Hydrodynamic radii of near-spherical objects. The friction coefficient $\xi^T = F_D/U_V$ of a simple-cubic array of particles has been used to determine the effective hydrodynamic radius $a$ in a suspending fluid with kinematic viscosity $\nu = 1/6$. The radius of the equal-volume sphere $a_V$ is also shown; the input radius $a_0$ defines the boundary surface of the object.

| $a_0$ | $a_V$ | $L$ | $\xi^T$ | $a$ |
|---|---|---|---|---|
| 1.5 | 1.66 | 6 | 344 | 1.53 |
|  |  | 8 | 242 | 1.54 |
|  |  | 16 | 159 | 1.54 |
| 2.5 | 2.68 | 10 | 596 | 2.60 |
|  |  | 16 | 354 | 2.61 |
|  |  | 32 | 256 | 2.61 |
| 4.5 | 4.53 | 16 | 1206 | 4.50 |
|  |  | 20 | 847 | 4.52 |
|  |  | 32 | 559 | 4.53 |
| 8.5 | 8.48 | 32 | 2015 | 8.47 |

TABLE III. Hydrodynamic radii of near-spherical objects as a function of fluid viscosity. The hydrodynamic radius of an object with input radius $a_0 = 4.5$ ($a_V = 4.53$) is shown for various values of the kinematic viscosity $\nu$ of the fluid. Numerical values for $a$ were determined for a simple cubic lattice as in Table II.

| $\nu$ | $a$ | $\nu$ | $a$ |
|---|---|---|---|
| 1/1536 | 5.12 | 1/6 | 4.53 |
| 1/96 | 4.79 | 1/3 | 4.42 |
| 1/24 | 4.67 | 2/3 | 4.27 |
| 1/12 | 4.60 | 8/3 | 3.70 |

TABLE IV. Steady flow past a column of cylinders up to Reynolds numbers $R_e = 100$ ($R_e = 2U_0 a/\nu$). Simulations (LBE) for cylinders of radii $a = 2.1$, $a = 5.1$, and $a = 10.1$ are compared with accurate finite difference solutions (FD) of the time-independent Navier-Stokes equations (Fornberg, 1991) for the same channel width (or cylinder separation) $W = 10a$. In addition to the drag coefficient $C_D = F_D/\rho U_0^2 a$, the wake length $L_W$, the minimum stream function $\Psi_c$, and its location $[X_c, Y_c]$, were determined from the contour plots (Fig. 6) of the recirculation zone behind the cylinder. At a Reynolds number $R_e = 100$ the flow is eventually unsteady; the Strouhal number $S = \omega_c a/U_0$ is also reported ($\omega_c$ is vortex shedding frequency).

| $R_e$ | $a$  | $C_D$ | $L_W$ | $\Psi_c$ | $X_c$ | $Y_c$ | $S$  |
|-------|------|-------|-------|----------|-------|-------|------|
| 10    | 2.1  | 3.76  |       |          |       |       |      |
|       | 5.1  | 4.08  | 1.6   |          |       |       |      |
|       | 10.1 | 4.21  | 1.5   | -0.0001  | 1.2   | 0.2   |      |
|       | FD   | 4.32  | 1.49  | -0.0006  | 1.22  | 0.23  |      |
| 20    | 2.1  | 2.60  |       |          |       |       |      |
|       | 5.1  | 2.82  | 2.5   | -0.02    | 1.7   | 0.4   |      |
|       | 10.1 | 2.91  | 2.6   | -0.009   | 1.7   | 0.4   |      |
|       | FD   | 2.98  | 2.65  | -0.011   | 1.65  | 0.40  |      |
| 40    | 2.1  | 1.97  |       |          |       |       |      |
|       | 5.1  | 2.11  | 4.7   | -0.03    | 2.0   | 0.5   |      |
|       | 10.1 | 2.17  | 4.7   | -0.039   | 2.31  | 0.53  |      |
|       | FD   | 2.19  | 4.74  | -0.042   | 2.24  | 0.54  |      |
| 100   | 5.1  | 1.70  | 9.0   | -0.11    | 3.7   | 0.6   | 0.11 |
|       | 10.1 | 1.67  | 10.7  | -0.121   | 3.91  | 0.79  | 0.10 |
|       | FD   | 1.61  | 10.3  |          |       |       |      |

TABLE V. Steady flow past a simple-cubic array of spheres at maximum packing $\epsilon = 1 - \pi/6$. The volume-averaged flow velocity $U_V$ (Eq. 3.4) was computed for various pressure gradients and sphere sizes. Results labeled by a drag force $F_D = 0$ correspond to creeping flow. The $R_e = 0$ friction coefficient (Zick and Homsy, 1982), $F_D/(6\pi\eta a U_V) = 42.1$, can be reproduced to within 2%. The Reynolds number $R_e = 2U_V a/\epsilon\nu$.

| $F_D$ | $L$ | $F_D/(6\pi\eta a U_V)$ | $R_e$ | $F_D$ | $L$ | $F_D/(6\pi\eta a U_V)$ | $R_e$ |
|---|---|---|---|---|---|---|---|
| 0 | 9 | 49.1 | 0 | 5 | 9 | 49.5 | 9.4 |
|  | 17 | 42.7 | 0 |  | 17 | 44.0 | 10.5 |
|  | 33 | 42.8 | 0 |  | 33 | 44.4 | 10.5 |
| 10 | 9 | 50.5 | 18.4 | 20 | 9 | 52.4 | 35.4 |
|  | 17 | 45.9 | 20.2 |  | 17 | 48.3 | 38.4 |
|  | 33 | 46.6 | 19.9 |  | 33 | 49.3 | 37.6 |
| 40 | 9 | 55.2 | 67.2 | 70 | 9 | 58.4 | 111.2 |
|  | 17 | 50.9 | 72.9 |  | 17 | 52.9 | 122.9 |
|  | 33 | 52.2 | 71.1 |  | 33 | 54.6 | 118.9 |
|  | 65 | 52.5 | 70.7 |  | 65 | 55.0 | 118.1 |
| 100 | 33 | 56.4 | 164.5 |  |  |  |  |
|  | 65 | 56.8 | 163.3 | 130 | 65 | 58.4 | 206.7 |

TABLE VI. Translational and rotational temperatures of suspended spheres, based on mean square translational and rotational velocities; the effective temperature of the fluctuating fluid is 0.25.

| $\phi$ | $a$ | $N = 16$ | | $N = 128$ | |
| --- | --- | --- | --- | --- | --- |
| | | $T_{tran}$ | $T_{rot}$ | $T_{tran}$ | $T_{rot}$ |
| 0.05 | 1.54 | 0.196 | 0.164 | 0.202 | 0.166 |
| | 2.61 | 0.215 | 0.202 | 0.218 | 0.201 |
| 0.25 | 1.54 | | | 0.183 | 0.158 |
| | 2.61 | 0.195 | 0.196 | 0.204 | 0.196 |
| | 4.53 | 0.202 | 0.213 | | |
| 0.45 | 2.61 | | | 0.183 | 0.180 |
| | 4.53 | 0.187 | 0.206 | 0.193 | 0.203 |
| | 8.47 | 0.207 | 0.226 | | |

# FIGURES

Fig. 1. Geometry for quasi-periodic simulations. The figure illustrates a system that closely approximates a periodic one under an external flow. The lattice nodes are shown by solid circles and the boundary nodes by solid squares. The arrows indicate velocity directions $\mathbf{c}_i$ and $\mathbf{c}'_i$ at the boundary nodes (c.f. Fig. 2 of paper I). Periodic boundary conditions are applied across the planes indicated by solid lines; the configuration of solid particles within the quasi-periodic unit cells (bounded by dashed lines) are identical. Macroscopic flows can be set up by the planes of boundary nodes at either end the the system. With this geometry we can set up uniform flow perpendicular to the boundary walls or an approximately linear shear flow parallel to the boundary walls. The properties of the central cell are close to those of a truly periodic system; it is not necessary, apparently, to include more cells, although this can be done if required.

Fig. 2. Translational and rotational friction coefficients of a simple-cubic lattice of spheres. The drag coefficients, normalized by the isolated sphere values, are plotted as a function of volume fraction for several different size objects. The solid lines are determined from accurate numerical solutions of the Stokes equations (Ladd, 1988; Ladd, 1990), as indicated in the text.

Fig. 3. Hydrodynamic interactions between pairs of spheres. The perpendicular and parallel friction coefficients are plotted as a function of $s = R/a - 2$. The results at $s = 0$ are for objects in closest possible proximity. The systems are periodic, with a two-sphere unit cell. The solid lines are again solutions of the Stokes equations in the same periodic geometry.

Fig. 4. Hydrodynamic transport coefficients for random arrays of spheres. Results from simulations of 16 spheres (with periodic boundary conditions) are compared with accurate numerical solutions of the Stokes equations (Ladd, 1990). The lattice-Boltzmann results (present work) are plotted as symbols; results from Ladd, 1990 are shown as solid lines. The statistical errors in both sets of calculations are smaller than the plotting symbols.

Fig. 5. Velocity profiles in sheared suspensions at 5% and 45% solids concentration. The circles show the ensemble-averaged velocity profile; the solid circles indicate the central cell from which the viscosities shown in Fig. 4 are calculated. The velocities are scaled by the shear rate $\gamma_0$ of a sample of pure fluid under the same conditions. In dense suspensions, there is a rapid change in effective viscosity from the pure fluid regions ($x < 0$ and $x > 3L$) to the suspension; thus the velocity profile is non-linear in this region. However the region surrounding the central cell has a linear velocity profile at all densities, although the local shear rate is less than $\gamma_0$.

Fig. 6. Stream lines for steady incompressible flow past a column of circular cylinders at various Reynold's numbers. A small portion (1/8) of the total channel length, near the cylinder, is shown.

Fig. 7. Pressure drop as a function of Reynolds number for steady flow past a simple-cubic array of spheres. The simulation results for different sized spheres are shown as solid symbols; the open symbols are the finite element results (Tompson, 1983). The solid line was fitted to the data of Lahbabi and Chang, 1985.

Fig. 8. Frequency-dependent torque on a rotating sphere. The modulus and phase lag (in degrees) of the drag torque on a sphere undergoing small amplitude rotational oscillations are shown as a function of the reduced frequency $\omega a^2/\nu$. Results are shown for two different size spheres at the same fluid viscosity $\nu = 1/6$; the solid lines are taken from expressions give in Landau and Lifshitz, 1959.

Fig. 9. Frequency-dependent torque on a rotating sphere as a function of the relative viscosity inside and outside the sphere. The modulus and phase lag (in degrees) of the drag torque on a sphere undergoing small amplitude rotational oscillations are shown as a function of the reduced frequency $\omega a^2/\nu_{ext}$, for various viscosity ratios $\nu_r = \nu_{int}/\nu_{ext}$. Results are shown for viscosity ratios of 4, and 16; the hydrodynamic radii of the spheres are (from Table III) 4.67, and 4.79 respectively. The dashed lines include the full interior torque, whereas the solid line includes just the inertial contribution, corresponding to $\nu_r = \infty$. Note that the reduced frequency is based on the exterior fluid viscosity $\nu_{ext}$.

Fig. 10. Translational velocity $U(t)$ and rotational velocity $\Omega(t)$ of an isolated sphere. The time-dependent velocities of the sphere are shown as solid symbols; the solid lines are theoretical results, obtained by an inverse Laplace transform of the frequency dependent friction coefficients (Hauge and Martin-Löf, 1973) of a sphere of appropriate size ($a = 2.61$) and effective mass ($\rho_s/\rho = 11$); the kinematic viscosity of the pure fluid $\nu = 1/6$.

Fig. 11. Translational velocity correlation function $\langle U(t)U(0)\rangle$ and rotational velocity correlation function $\langle \Omega(t)\Omega(0)\rangle$ of an isolated sphere, normalized to their $t = 0$ values. The time-dependent correlation functions are shown as solid symbols (with statistical error bars). The solid lines are theoretical results, obtained by an inverse Laplace transform of the frequency dependent friction coefficients (Hauge and Martin-Löf, 1973) of a sphere of appropriate size ($a = 2.61$) and mass ($\rho_s/\rho = 11$); the kinematic viscosity of the pure fluid $\nu = 1/6$.

Fig. 12. Scaled mean-square displacement $\langle \Delta R^2(t)\rangle/6D_s t$ at short times, vs. reduced time $t/\tau$. Simulation results for 128 spheres (solid symbols) are shown at packing fractions $\phi$ of 5%, 25% and 45%; the solid line is the isolated sphere result. The simulation parameters were the same as in Fig. 11 except that a sphere of radius 4.5 was used at the highest volume fraction.

Fig. 13. Scaled relaxation time $\tau/\tau_0$ ($\tau_0 = a^2/\nu$) and self-diffusion coefficient $D_s/D_0$ vs. packing fraction $\phi$. The data points are determined from the scaling of the 128-sphere simulation data for various particle sizes. The uncertainty in fitting the data is about 5%. Independent results (Ladd, 1990) for $\eta_0/\eta$ (solid lines) and $D_s/D_0$ (dashed lines) are shown for comparison.

Fig. 14. Diffusion coefficients for random arrays of spheres. Self-diffusion coefficient $D_s$ and collective mobility $\mu$ from simulations of 16 spheres (filled symbols) and 128 spheres (open symbols) are compared with numerical solutions of the Stokes equations (Ladd, 1990). Results from moment expansions of the Oseen equation are shown as solid lines ($N = 16$) and dashed lines ($N = 128$). The statistical errors in the fluctuating lattice-Boltzmann simulations are comparable to the size of the plotting symbols.

Fig. 15. High-frequency shear viscosity for random arrays of spheres. Shear viscosities from simulations of 16 spheres (filled symbols) and 128 spheres (open symbols) are compared with numerical solutions of the Stokes equations (Ladd, 1990). The statistical errors in the fluctuating lattice-Boltzmann simulations are comparable to the size of the plotting symbols.

# APPENDIX: COLLECTIVE DIFFUSION COEFFICIENT

The Green-Kubo expression for the collective diffusion coefficient of a binary mixture can be derived from the fluctuations in concentration; here we outline the standard procedure given, for example, in Hansen and McDonald, 1986. The macroscopic equations of continuity for the two species,

$$\partial_t \rho_i + \boldsymbol{\nabla} \cdot (\rho_i \mathbf{u}_i) = 0, \tag{A1}$$

can be cast into the form of an equation for conservation of mass ($\rho = \rho_1 + \rho_2$) of the usual form, and an equation for the (mass weighted) concentration $c = \rho_1/\rho$

$$\partial_t c + \mathbf{u} \cdot \boldsymbol{\nabla} c = -\rho^{-1} \boldsymbol{\nabla} \cdot \mathbf{j}; \tag{A2}$$

$\mathbf{j} = \mathbf{j}_1 = \rho_1(\mathbf{u}_1 - \mathbf{u})$ is the diffusive flux of species 1, defined relative to the local stream velocity $\mathbf{u} = (\rho_1 \mathbf{u}_1 + \rho_2 \mathbf{u}_2)/\rho$. The diffusive flux can also be written in a more symmetric form $\mathbf{j} = (1-c)\rho_1 \mathbf{u}_1 - c\rho_2 \mathbf{u}_2$.

A microscopic expression (in terms of molecular coordinates) for the spatial Fourier transform of the concentration can derived by linearizing the fluctuations in $c$ about the mean densities $\overline{\rho_i}$,

$$\delta c(\mathbf{k}, t) = \frac{\overline{\rho_2} \delta \rho_1(\mathbf{k}, t) - \overline{\rho_1} \delta \rho_2(\mathbf{k}, t)}{\overline{\rho}^2} = \frac{X_2 \sum_{i \epsilon 1} e^{-i\mathbf{k} \cdot \mathbf{R}_i} M_1 - X_1 \sum_{i \epsilon 2} e^{-i\mathbf{k} \cdot \mathbf{R}_i} M_2}{\overline{\rho}}; \tag{A3}$$

here $X_i$ indicate average concentrations $\overline{\rho_i}/\overline{\rho}$, that are spatially invariant. Similarly, the diffusive flux can be expressed microscopically as

$$\mathbf{j}(\mathbf{k}, t) = \int e^{-i\mathbf{k} \cdot \mathbf{r}} \mathbf{j}(\mathbf{r}, t) d\mathbf{r} = X_2 \sum_{i \epsilon 1} e^{-i\mathbf{k} \cdot \mathbf{R}_i} M_1 \mathbf{U_i} - X_1 \sum_{i \epsilon 2} e^{-i\mathbf{k} \cdot \mathbf{R}_i} M_2 \mathbf{U_i}. \tag{A4}$$

Using the constitutive equation

$$\mathbf{j} = -\rho D \boldsymbol{\nabla} c \tag{A5}$$

to define the collective diffusion coefficient $D$, we find after some elementary manipulations that

$$D = \frac{\int_0^\infty \langle \mathbf{J}(t) \cdot \mathbf{J}(0) \rangle \, dt}{3\rho^2 \langle c(\mathbf{k})c(-\mathbf{k}) \rangle_{\mathbf{k}=0}} \tag{A6}$$

where $\mathbf{J}$ is the $k = 0$ limit of Eq. A4 and the small $k$ limit is taken in the denominator also. The equal-time concentration fluctuations can be expressed in terms of partial structure factors

$$S_{ab}(\mathbf{k}) = \frac{1}{\sqrt{N_a N_b}} \sum_{i \epsilon a} \sum_{i \epsilon b} e^{-i\mathbf{k} \cdot \mathbf{R}_{ij}}; \tag{A7}$$

$$\rho^2 \langle c(\mathbf{k})c(-\mathbf{k}) \rangle_{\mathbf{k}=0} = X_1 X_2 M_1 M_2 \left[ N_2 S_{11}(0) - 2\sqrt{N_1 N_2} S_{12}(0) + N_1 S_{22}(0) \right]. \tag{A8}$$

The equal-time fluctuations in the diffusive flux are readily evaluated from the Maxwell-Boltzmann distribution of velocities,

$$\langle \mathbf{J}(0) \cdot \mathbf{J}(0) \rangle = 3k_B T X_1 X_2 (N_1 M_1 + N_2 M_2). \tag{A9}$$

We now specialize to the case of a particle suspension, for which the number of solid spheres ($N_1$) is much less than the number of solvent molecules ($N_2$); thus Eq. A8 simplifies to

$$\rho^2 \langle c(\mathbf{k})c(-\mathbf{k}) \rangle_{\mathbf{k}=0} = X_1 X_2 M_1 M_2 N_2 S_{11}(0), \tag{A10}$$

where $S_{11}(\mathbf{k})$ is the structure factor for the solid particles. From Eqs. A6, A9, and A10,

$$D = \frac{k_B T}{M_1 X_2 S_{11}(0)} \int_0^\infty \frac{\langle \mathbf{J}(t) \cdot \mathbf{J}(0) \rangle}{\langle \mathbf{J}(0) \cdot \mathbf{J}(0) \rangle} dt. \tag{A11}$$

In the simulations, fluctuations in $\mathbf{J}$ can be related to fluctuations in the velocity of the solid phase (by momentum conservation); i.e. $\mathbf{J} = M_1 \sum_{i \epsilon 1} \mathbf{U}_i = M_1 \mathbf{U}_c$ (Eq. 6.6). Using the relation

$$\mu = \frac{S_{11}(0) D}{k_B T} \tag{A12}$$

we arrive at Eq. 6.7 for the ratio $\mu/\mu_0$.

The factor $X_2^{-1}$ arises from the conservation of total momentum, which is implied in the definition of concentration fluctuations (Eq. A3). In a canonical ensemble, where the total momentum and energy are free to fluctuate,

$$\langle \mathbf{U}_c(0) \cdot \mathbf{U}_c(0) \rangle = \frac{3 N_1 k_B T}{M_1}. \tag{A13}$$

On the other hand, at fixed total momentum $\mathbf{P}$, fluctuations in $\mathbf{U}_c$ are reduced (Lebowitz et al., 1967),

$$\langle \mathbf{U}_c(0) \cdot \mathbf{U}_c(0) \rangle_{\mathbf{P}=0} = \langle \mathbf{U}_c(0) \cdot \mathbf{U}_c(0) \rangle - \langle \mathbf{PP} \rangle : \partial_{\mathbf{P}} \langle \mathbf{U}_c \rangle_{\mathbf{P}} \cdot \partial_{\mathbf{P}} \langle \mathbf{U}_c \rangle_{\mathbf{P}}$$
$$= \langle \mathbf{U}_c(0) \cdot \mathbf{U}_c(0) \rangle - \mathcal{M} k_B T \partial_{\mathbf{P}} \langle \mathbf{U}_c \rangle_{\mathbf{P}} : \partial_{\mathbf{P}} \langle \mathbf{U}_c \rangle_{\mathbf{P}}, \tag{A14}$$

where $\mathcal{M} = N_1 M_1 + N_2 M_2$ is the total mass of solid and fluid. For fixed total momentum, the average particle velocity is $\mathbf{P}/\mathcal{M}$, and

$$\partial_{\mathbf{P}} \langle \mathbf{U}_c \rangle_{\mathbf{P}} = \frac{N_1}{\mathcal{M}} \mathbf{1}. \tag{A15}$$

Combining Eqs. A16 and A15, the fluctuations at constant total momentum are found to be

$$\langle \mathbf{U}_c(0) \cdot \mathbf{U}_c(0) \rangle_{\mathbf{P}=0} = \frac{3 N_1 k_B T X_2}{M_1}. \tag{A16}$$

Thus the ensemble dependence of the zero-time fluctuations accounts for the factor $X_2^{-1}$; in an ensemble where the total momentum is free to fluctuate this factor is absent.